\documentclass[article]{aa}  
\usepackage{txfonts}
\usepackage{natbib}
\usepackage{graphicx}
\usepackage{color}
\usepackage{longtable,lscape}
\usepackage{float}
\usepackage{subfigure}
\newcommand{\kmpers}{$\mathrm{km \, s^{-1}}$} 

\newcommand{\cmtwo}{cm$^{-2}$} 
 
\newcommand{\cmthree}{cm$^{-3}$}

\newcommand{\vlsr}{$\upsilon_{\rm LSR}$}              

\newcommand{\tmb}{$T_{\rm mb}$}

\newcommand{\about}{$\sim$}                       
\newcommand{\expo}[1]{$10^{#1}$}
\newcommand{\texpo}[1]{$\,\times\,10^{#1}$}

\newcommand{\amindot}[2]{\mbox{#1$\stackrel {\prime}{_{\bf \cdot}}$#2}}

\newcommand{\nhtva}{$n$($\mathrm{H_2}$)}           

\newcommand{\htvao}{H$_2$O}  
\newcommand{\htva}{H$_2$}

\newcommand{\oishort}{[O\,{\sc i}]\,63\,$\mu$m}
\newcommand{\oilong}{[O\,{\sc i}]\,145\,$\mu$m}

\newcommand{\cotretva}{CO\,(3$-$2)}

\newcommand{\cotionio}{CO\,(10$-$9)}
\newcommand{\cofemtonfjorton}{CO\,(15$-$14)}
\newcommand{\asec}{$^{\prime \prime}$}
\newcommand{\adeg}{$^{\circ}$}

\newcommand{\atwozero}{$\alpha_{2000}$}
\newcommand{\dtwozero}{$\delta_{2000}$}

\newcommand{\radot}[4]{\mbox{#1$^{\rm h}$#2$^{\rm m}$#3$\stackrel{^{\rm
s}}{_{\bf\cdot}}$#4}}
\newcommand{\decdot}[3]{\mbox{#1$^{\circ}$#2$^{\prime}$#3$^{\prime \prime}$}}
\newcommand{\herschel}{\textit{Herschel}}
\newcommand{\hifi}{{HIFI}}
\newcommand{\pacs}{{PACS}}
\newcommand{\spire}{{SPIRE}}

\newcommand{\spitzer}{\textit{Spitzer}}
\newcommand{\vlarhooph}{\mbox{VLA\,1623}}

\titlerunning{Resolving the shocked gas in HH\,54 with \herschel}
\authorrunning{P. Bjerkeli et al.}

\begin{document}
   \title{Resolving the shocked gas in HH\,54 with \herschel\thanks{\herschel\ is an ESA space observatory with science instruments provided by European-led Principal Investigator consortia and with important participation from NASA.}}

   \subtitle{CO line mapping at high spatial and spectral resolution}

   \author{Bjerkeli, P.
          \inst{1,2,3}
          \and
          Liseau, R.\inst{3}
          \and
          Brinch, C.\inst{1,2}
          \and
          Olofsson, G.\inst{4}
          \and
          Santangelo, G.\inst{5,6}
          \and
          Cabrit, S.\inst{7}
	\and 
	Benedettini, M.\inst{5}
	\and \\
	Black, J.~H.\inst{3}
	\and 
	Herczeg, G.\inst{8}
	\and
	Justtanont, K.\inst{3}
	\and
	Kristensen, L. E.\inst{9}
	\and
	Larsson, B.\inst{4}
	 \and
          Nisini, B.\inst{6}
	\and
	Tafalla, M.\inst{10}
          }
   \institute{
             Niels Bohr Institute, University of Copenhagen, Juliane Maries Vej 30, DK-2100 Copenhagen \O, Denmark \\
             \email{per.bjerkeli@nbi.dk}
             \and
             Centre for Star and Planet Formation and Natural History Museum of Denmark, University of Copenhagen, \O ster Voldgade 5--7, DK-1350 Copenhagen K, Denmark 
             \and
             Department of Earth and Space Sciences, Chalmers University of Technology, Onsala Space Observatory, 439 92 Onsala, Sweden 
  \and
Department of Astronomy, Stockholm University, AlbaNova, 106 91 Stockholm, Sweden
\and
INAF - Osservatorio Astrofisico di Arcetri, Largo E. Fermi 5, 50125 Firenze, Italy
\and
INAF - Osservatorio Astronomico di Roma, Via di Frascati 33, 00040 Monte Porzio Catone, Italy
\and
LERMA, Observatoire de Paris, UMR CNRS 8112, 61 Av. de l'Observatoire, 75014 Paris, France
\and
Kavli Institute for Astronomy and Astrophysics, Peking University, Beijing, 100871, PR China
\and
Harvard-Smithsonian Center for Astrophysics, 60 Garden Street, Cambridge, MA 02138, USA
\and
Observatorio Astron—mico Nacional (IGN), Alfonso XII 3, 28014 Madrid, Spain
                          }

   \date{Received 11 August 2014 / Accepted 22 September 2014}

  \abstract
   {The HH\,54 shock is a Herbig-Haro object, located in the nearby Chamaeleon $\tt{II}$ cloud. Observed CO line profiles are due to a complex distribution in density, temperature, velocity, and geometry.}
   {Resolving the HH\,54 shock wave in the far-infrared cooling lines of CO constrain the kinematics, morphology, and physical conditions of the shocked region.}
   {We used the PACS and SPIRE instruments on board the \herschel\ space observatory to map the full FIR spectrum in a region covering the HH\,54 shock wave. Complementary \herschel-HIFI, APEX, and \spitzer\ data are used in the analysis as well. The observed features in the line profiles are reproduced using a 3D radiative transfer model of a bow-shock, constructed with the Line Modeling Engine code (LIME).}
   {The FIR emission is confined to the HH\,54 region and a coherent displacement of the location of the emission maximum of CO with increasing $J$ is observed. The peak positions of the high-$J$ CO lines are shifted upstream from the lower $J$ CO lines and coincide with the position of the spectral feature identified previously in \cotionio\ profiles with HIFI. This indicates a hotter molecular component in the upstream gas with distinct dynamics. The coherent displacement with increasing $J$ for CO is consistent with a scenario where IRAS12500 -- 7658 is the exciting source of the flow, and the 180~K bow-shock is accompanied by a hot (800~K) molecular component located upstream from the apex of the shock and blueshifted by $-$7~\kmpers. The spatial proximity of this knot to the peaks of the atomic fine-structure emission lines observed with \spitzer\ and PACS ([O\,{\sc i}]63, 145$\mu$m) suggests that it may be associated with the dissociative shock as the jet impacts slower moving gas in the HH\,54 bow-shock.}
   {}


   \maketitle
%
\section{Introduction}
Herbig-Haro objects \citep[see e.g.][]{Herbig:1950kx,Haro:1952fk} are the optical manifestations of molecular outflows \citep{Liseau:1986uq} emanating from young stellar objects (YSOs).  They reveal the shocked gas moving at the highest velocities. 

A shock where the far-infrared cooling lines can be studied in detail using a very limited amount of \herschel\ \citep{Pilbratt:2010kx} time is HH\,54, located in the Chamaeleon~II cloud. At a distance of only 180~pc \citep{Whittet:1997kx} it is close enough to be spatially resolved with \herschel. The source of the HH\,54 flow is, however, a matter of debate. Several nearby IRAS point sources have been proposed as the exciting source \citep[see e.g.][and references therein]{Caratti-o-Garatti:2009fk,Bjerkeli:2011qy}, but the issue remains essentially unsettled. Previous observations have shown that observed CO line profiles, from $J_{\rm{up}}$~=~2 to 10, are due to a complex distribution in temperature, density, velocity, and geometry \citep{Bjerkeli:2011qy}. In particular, these observations revealed the presence of a spectral bump, tracing a distinctly different gas component than the warm gas (\about200~K) responsible for the bulk part of the CO line emission.  

In this paper, we present spectroscopic observations (50 -- 670 $\mu$m) of CO at a high spatial resolution acquired with PACS \citep{Poglitsch:2010uq} and SPIRE \citep{Griffin:2010lr}. We also present velocity resolved observations of the \cofemtonfjorton\ line and the CO\,(3--2) line obtained with HIFI and APEX, respectively. These datasets are analysed in conjunction with previously published CO and \htva\ data acquired with \herschel\ and \spitzer, providing valuable information on the kinematics, morphology, and physical conditions in the shocked region. 
\section{Observations}
The OT2 \herschel\ data\footnote{{Obs. IDs: 1342248309, 1342248543, 1342257933, and 1342250484}} presented here were observed in \pacs\ 
range spectroscopy mode in July 2012, SPIRE point spectroscopy mode in December 2012, and HIFI single pointing mode in September 2012. The observations were centred on HH\,54~B, \atwozero~=~\radot{12}{55}{50}{3}, \dtwozero~=~\decdot{--76}{56}{23} \citep{Sandell:1987vn},  and covered the wavelength region  \about 50 -- 670 $\mu$m. The \pacs\ receiver includes 5 by 5 spatial pixels (spaxels), each covering a region of 9.4\asec\ by 9.4\asec. The \spire\ spectrometer has a field of view of \amindot{2}{6} and full spatial sampling was used to allow the determination of the position of the emission maximum to a high accuracy. In this observing mode, the Beam Steering Mirror is moved in a 16-point jiggle to provide complete Nyquist sampling of the observed region. In both the \pacs\ and \spire\ observations, each emission line is unresolved in velocity space. The velocity resolved spectrum of the \cofemtonfjorton\ line was observed with \hifi\ \citep{de-Graauw:2010uq}, where the full width half maximum (FWHM) of the beam is 12\asec. The channel spacing is 500 kHz (0.09 \kmpers\ at 1727~GHz) and the spectrum was converted to the \tmb\ scale using a main-beam efficiency of 0.76.
The data were reduced in IDL and Matlab.

The \cotretva\ observations, where the FWHM of the beam is 18\asec, were acquired with the APEX-2 receiver\footnote{http://www.apex-telescope.org/heterodyne/shfi/het345/} in March 2012. A rectangular region \mbox{(80\asec\ $\times$ 80\asec)} centred on HH\,54~B was observed and the channel spacing is 76 kHz (0.06 \kmpers\ at 346~GHz). The data reduction was performed in Matlab and the spectrum was converted to the $T_{\rm{mb}}$ scale using the main beam efficiency 0.73.

For the analysis, we also make use of already published CO data \citep{Bjerkeli:2011qy} and \htva\ data \citep{Neufeld:2006fk}. For the observational details we refer the interested reader to those publications. 
   
\section{Results}
\begin{figure*}[]
   \centering
   \rotatebox{90}{\includegraphics[width=1.22\textwidth]{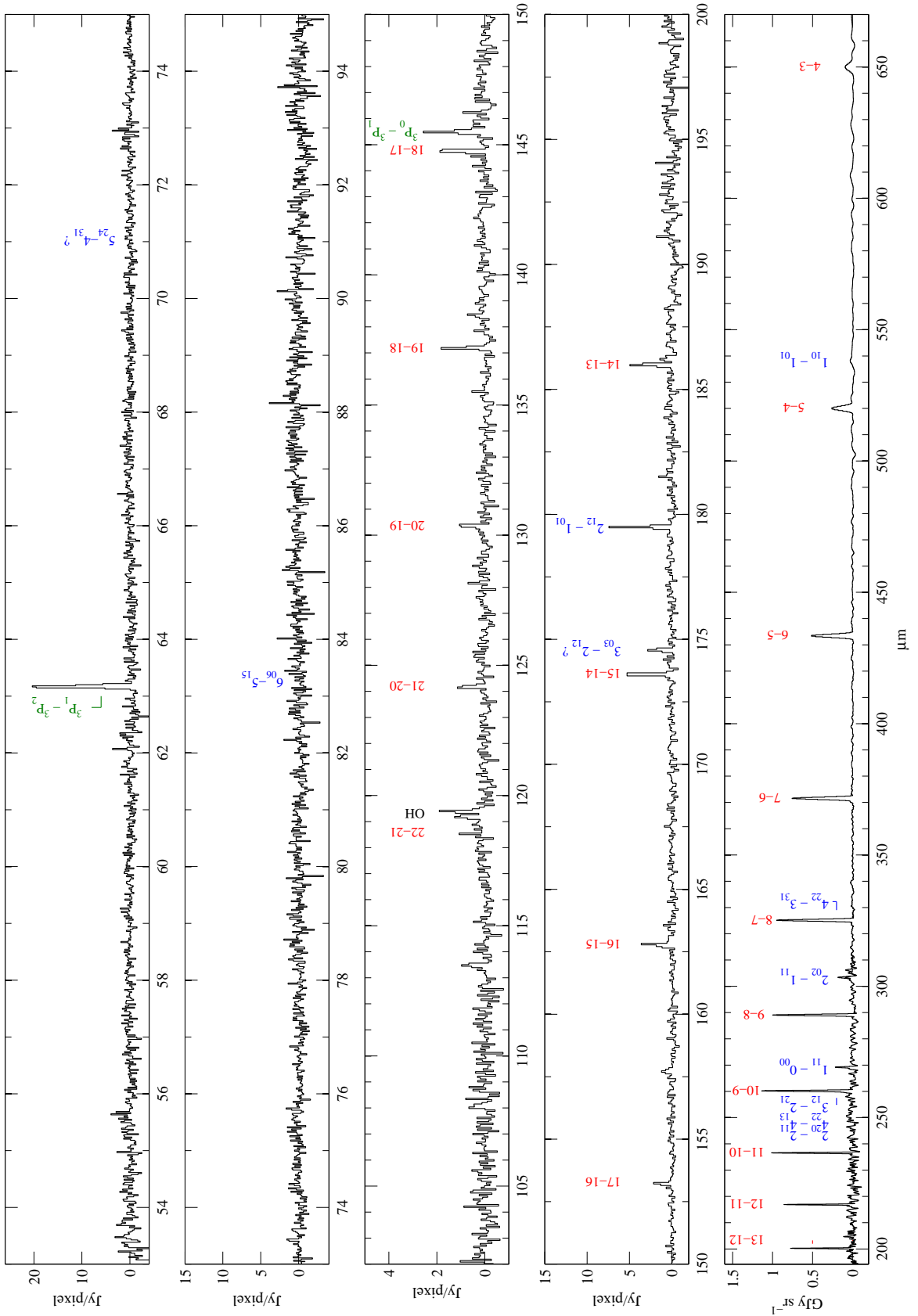}}
   
      \caption{\textit{Upper four panels:} PACS spectrum towards the central spaxel. Only lines that are detected with a signal-to-noise ratio greater than 3 somewhere in the mapped region (not necessarily the central spaxel) are marked as a detected line. Only upper and lower state energy levels are indicated for clarity. \textit{Lower panel:} Same as the other panels but for the SPIRE spectrum towards the central region. The [O\,{\sc{i}}] lines are marked in green, the \htvao\ lines in blue and the CO lines in red. In this figure, there are different units for the \pacs\ and \spire\ observations and no continuum subtraction has been made.}
         \label{fig:spectrum}
   \end{figure*}

\begin{figure}[ht]
   \centering
       \includegraphics[width=0.50\textwidth]{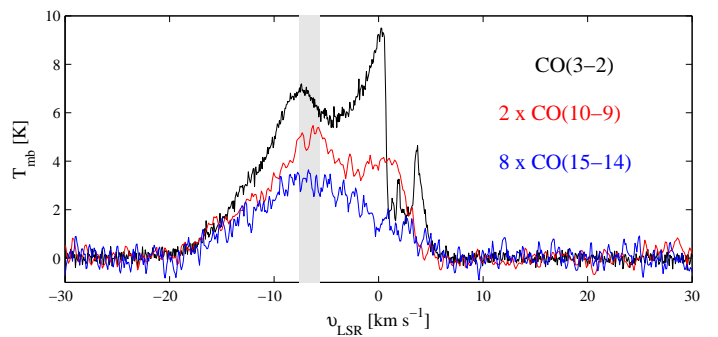}  
      \caption{Comparison between the \cotretva, \cotionio, and \cofemtonfjorton\ spectra. The latter two have been scaled for clarity. The peak velocity of the bump is within 2~\kmpers\ for all three lines (shaded area).}
         \label{fig:speccomp}
   \end{figure}

\begin{figure*}[ht]
   \centering
   
   \begin{tabular}{ll}
      \includegraphics[width=0.52\textwidth]{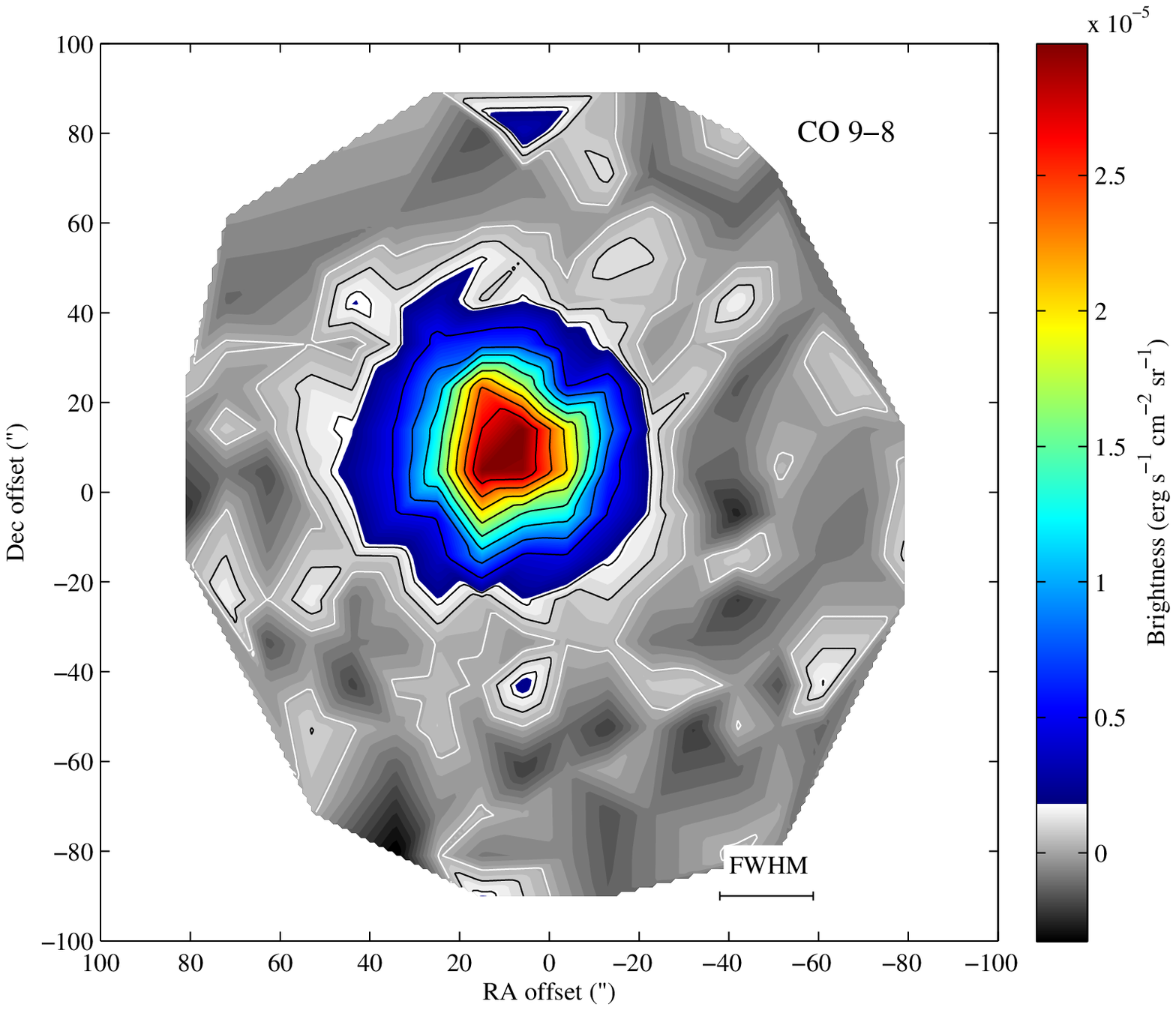}   & 
        \includegraphics[width=0.434\textwidth]{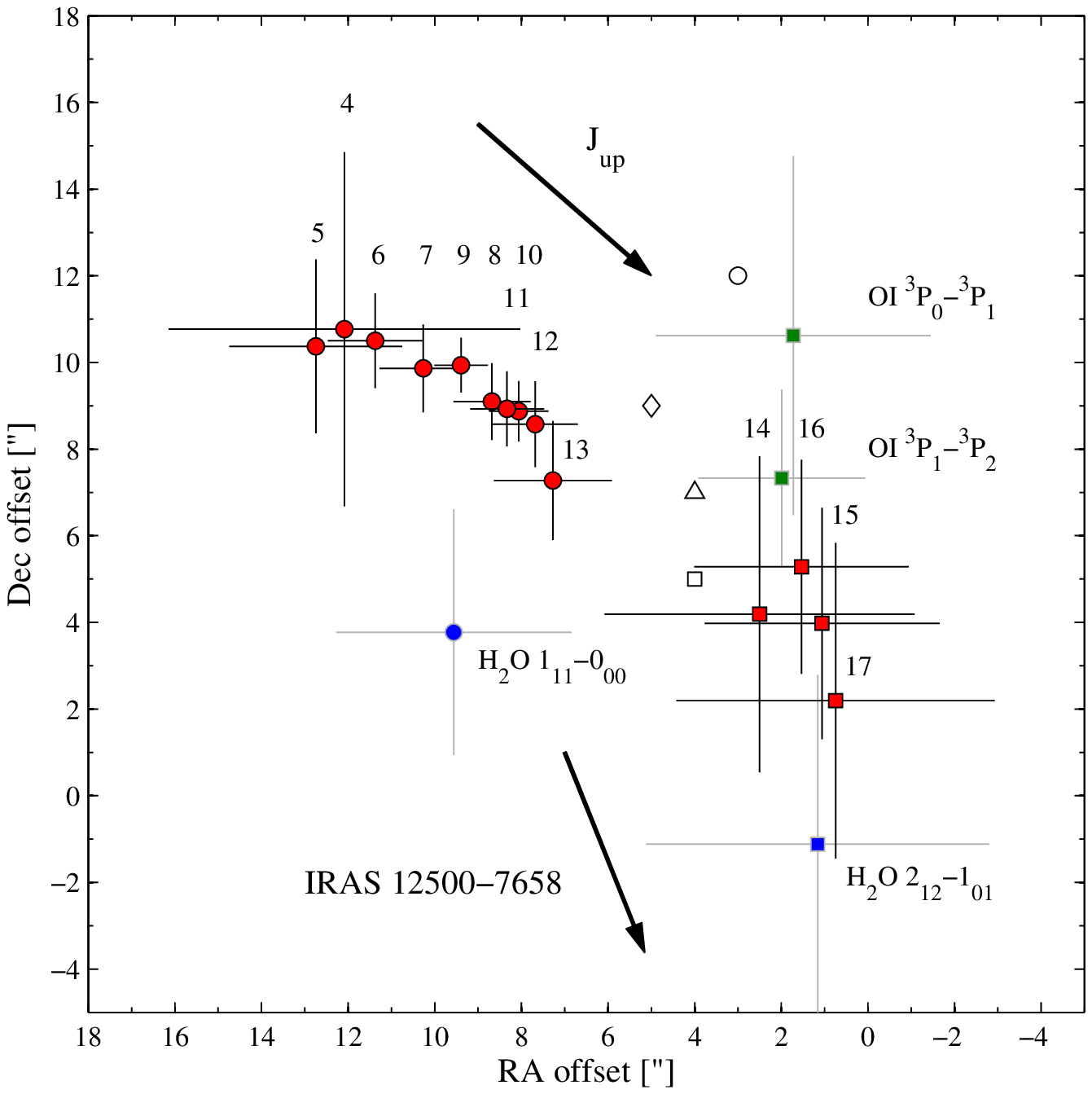} \\
   
      \end{tabular}
      \caption{\textit{Left panel:} The CO\,(9--8) line emission obtained with SPIRE. The coloured contours show where the signal-to-noise ratio is higher than 5, the black line indicate the 3$\sigma$ level and the white line the 1$\sigma$ level. The FWHM of the \spire\ beam is indicated in the lower right corner of this figure. \textit{Right panel:} The red (CO), blue (\htvao) and green ([O\,{\sc{i}}]) markers show the location of the emission maximum with error bars from the Gaussian fit. The open square, circle, triangle and diamond show the peak positions of the fine-structure emission lines [Ne\,{\sc ii}] 12.8~$\mu$m, [Fe\,{\sc ii}] 26~$\mu$m, [S\,{\sc i}] 25~$\mu$m, and [Si\,{\sc ii}] 35~$\mu$m lines, respectively \citep[][Fig.~3]{Neufeld:2006fk}. The arrows indicate the direction of increasing $J_{\rm{up}}$ for CO and the direction to IRAS\,12500--7658. The quantum numbers (only $J_{\rm{up}}$ for CO) are also shown in the figure. Angular offsets are given relative to the position of HH\,54 B.}
         \label{fig:gaussianfits}
   \end{figure*}

\begin{table}[t!]
    \caption{Total CO line fluxes and FWHM from the Gaussian fitting.}
      \label{table:results}
      \renewcommand{\footnoterule}{} 
      \begin{tabular}{lllllrccr}
        \noalign{\smallskip}
      \hline
      \hline
      \noalign{\smallskip}
       \noalign{\smallskip}
      Transition  & Frequency & Line flux & FWHM\\
      \noalign{\smallskip}
      & (GHz) & ($10^{-13}$ erg $\rm{s}^{-1}$ \cmtwo) & (\asec)  \\
      \noalign{\smallskip}
       \hline
        \noalign{\smallskip}
        \noalign{\smallskip}
      4 -- 3 	&  	461.041 		& 1.49 $\pm$ 0.77  & 52.8 $\pm$ 6.9\\
      5 -- 4 	&	576.268		& 4.21 $\pm$ 1.07  & 52.8 $\pm$ 3.4\\
      6 -- 5 	&	691.473 		&6.20 $\pm$	0.99	 & 45.6 $\pm$ 1.8\\		
      7 -- 6 	& 	806.652 		&9.84 $\pm$	1.41	 & 47.3 $\pm$ 1.7\\
      8 -- 7 	&	921.800 		&11.3 $\pm$	1.47		& 45.3 $\pm$ 1.5\\	
      9 -- 8	&	1036.912 		&8.71 $\pm$	1.20	        & 31.3 $\pm$ 1.0\\
      10 -- 9	&	1151.985 		&8.36 $\pm$	1.24		& 30.8 $\pm$ 1.1\\
      11 -- 10	&	1267.014 		&7.05 $\pm$	1.33	& 30.1 $\pm$ 1.4\\
      12 -- 11	&	1381.995 		&6.36 $\pm$	1.34	& 30.8 $\pm$ 1.6\\
      13 -- 12	&	1496.923 		&4.76 $\pm$	1.44	& 28.8 $\pm$ 2.1\\
      14 -- 13	&	1611.794 		&5.07 $\pm$	4.80	& 26.3 $\pm$ 6.5\\
      15 -- 14	&	1726.603 		&4.55 $\pm$	3.88	& 21.0 $\pm$ 4.5\\
      16 -- 15	&	1841.346 		&4.03 $\pm$	2.99	& 22.5 $\pm$ 4.3\\
      17 -- 16	&	1956.018 		&2.76 $\pm$	2.95	& 23.1 $\pm$ 6.2\\   
      \noalign{\smallskip}
      \noalign{\smallskip}
      \hline
    \end{tabular}
\end{table}
Figure~\ref{fig:spectrum} shows the spectrum obtained with PACS and SPIRE towards the position closest to HH\,54 B, i.e. the central spaxel for the PACS observations. 
No significant continuum emission is detected, however, all CO rotational emission lines from \mbox{$J_{\rm{up}} = 4$ to 22} are detected at the 5 $\sigma$ level (line flux) or higher in at least one position in the observed region. The CO line fluxes are consistent with the fluxes measured with ISO \citep{Nisini:1996fk,Giannini:2006lr}, within the error bars. The total line fluxes are calculated by fitting Gaussians to each emission map (see Fig.~\ref{fig:gaussianfits} and Sec.~\ref{sec:discussion}) and are presented in Table~\ref{table:results}. The higher-$J$ CO lines ($J_{\rm{up}}>$~17) are excluded from this table due to the few positions where a 5 $\sigma$ detection is reached. In addition to the CO ladder, one OH line, the [O\,{\sc i}] lines at 63 and 145 $\mu$m, and several \htvao\ lines are detected. The water emission from HH\,54 is further discussed in a forthcoming paper \citep{Santangelo:2014hl}. [O\,{\sc iii}] is not detected, nor is [N\,{\sc ii}] or [C\,{\sc ii}], however, there is a feature (158 $\mu$m) at the 2.5 $\sigma$ level. The emission from all lines is confined to the HH\,54 region. A bump-like feature \citep[also discussed in][]{Bjerkeli:2011qy} is clearly detected in the \cotretva\ spectrum and likely also in the \cofemtonfjorton\ spectrum. In Fig.~\ref{fig:speccomp}, these two lines are shown, compared to the \cotionio\ spectrum \citep{Bjerkeli:2011qy}. The blue-shifted extent is the same for all three lines, and the peak velocity of the bump is within 2~\kmpers\ for the three lines. The \cotretva\ observations suffered from contamination close to the cloud LSR velocity. This is of minor importance, however, for the analysis presented in this paper. 
\section{Discussion}
\label{sec:discussion}

\subsection{Structure of the shock}
\label{section:structureoftheshock}
Even though the observed lines are only marginally resolved spatially, we can still (thanks to the high signal-to-noise ratio) determine the position of each emission maximum to a high accuracy. We therefore fit 2D Gaussians to each CO map and measure the position of the maximum. The result is a coherent displacement with increasing $J$ (see Fig.~\ref{fig:gaussianfits}). In this figure, the peak position of the two [O\,{\sc i}] lines and the two \htvao\ lines that are detected at a sufficient signal-to-noise ratio are also plotted. The \mbox{CO\,(9--8)} map obtained with SPIRE is shown as well, as an illustrative example. Since the lines are observed simultaneously with each instrument, the measurement of the spatial shift as a function of $J$ is robust. One explanation to  this shift, since the shock is oriented almost along the line of sight \citep{Caratti-o-Garatti:2009fk}, could be different portions of the shocked surface having different temperatures and/or densities. A hot molecular knot (traced by high-$J$ CO) on this surface would always appear projected upstream with respect to the bow outer rim (i.e. with respect to the region where the low-$J$ CO lines peak). The observed gradient in excitation is, however, consistent with the scenario that was presented in \citet{Bjerkeli:2011qy}, i.e. the centroids of the high-$J$ CO lines coincide spatially with the bump-like feature detected in the \cotionio\ line, discussed in that paper (their Fig.~5). These authors suggested that the bump traces an unresolved hot, low-density component and this is likely the cause of the observed displacement with increasing $J$ (Sec.~\ref{section:observedlineprofiles}). If IRAS 12500-7658 is the exciting source, this component appears located upstream of the bow-shock in projection. The opposite trend is observed in other sources \citep[see e.g.][]{Santangelo:2014xy}. On the other hand, the situation is morphologically reminiscent to the L1157-B1 case \citep{Benedettini:2012zr}. The high-$J$ CO emission peak is also in that object shifted upstream by about 10\asec\ from the outer bow-shock rim delineated by lower excitation lines. In both sources, this high-$J$ CO seems spatially associated with ionic fine structure lines such as [Fe\,{\sc ii}] and [Ne\,{\sc ii}] \citep{Neufeld:2006fk} and [O\,{\sc i}] upstream of the bow-shock (see Fig.~\ref{fig:gaussianfits}). The bright and compact ionic emission in HH\,54 and L1157-B1 is understood as tracing the dissociative and ionizing shock, where the jet suddenly encounters the slow-moving ambient shell \citep{Neufeld:2006fk,Benedettini:2012zr}. Thus, the possibility exists that the high-$J$ CO emission, and the associated bump spectral feature, are produced in a region related to the reverse shock \citep[e.g.][]{Fridlund:1998lr}. The apparent association of high-$J$ CO emission with a strongly dissociative shock, both in HH\,54 and in L\,1157-B1, would imply that either the jet material is already partly molecular, or that molecule reformation is efficient. Testing these interpretations clearly calls for further modelling work, that lies outside the scope of the present paper. 
\subsection{Observed line profiles}
\label{section:observedlineprofiles}
To check whether a low density, hot component can give rise also to the observed bump in the line profiles, we have updated the model presented in \citet{Bjerkeli:2011qy}. The models discussed there, were limited by the 1D+ geometry of the ALI code, used for the calculation of the radiative transfer. In this paper, we instead use the accelerated Monte-Carlo code LIME \citep{Brinch:2010uq}, that allows the full 3D radiative transfer to be calculated. The LIME code  does not put any constraints on the complexity of the models that can be constructed. Nevertheless, we have chosen to keep it simple for clarity. We therefore, again, compute the radiative transfer for a model that is observed from the front, but in this case with a small inclination angle with respect to the line of sight, i.e. 25\adeg\ \citep{Caratti-o-Garatti:2009fk}. We use the density (\nhtva~= \expo{5}~\cmthree) and temperature ($T_{\rm{kin}}~=~180~\rm{K}$) inferred from the $\chi^2$ analysis presented in \citet[][their Sec. 4.3]{Bjerkeli:2011qy}. The difference between the model presented in that paper and the model presented here, is mainly the presence of a slightly under-dense (\nhtva~$\simeq$~\expo{4}~\cmthree), hot ($T_{\rm{kin}}$~$\simeq$~800~K) component that is located upstream of the bow-shock. 
 \begin{figure}[]
   \centering
     \includegraphics[width=0.48\textwidth]{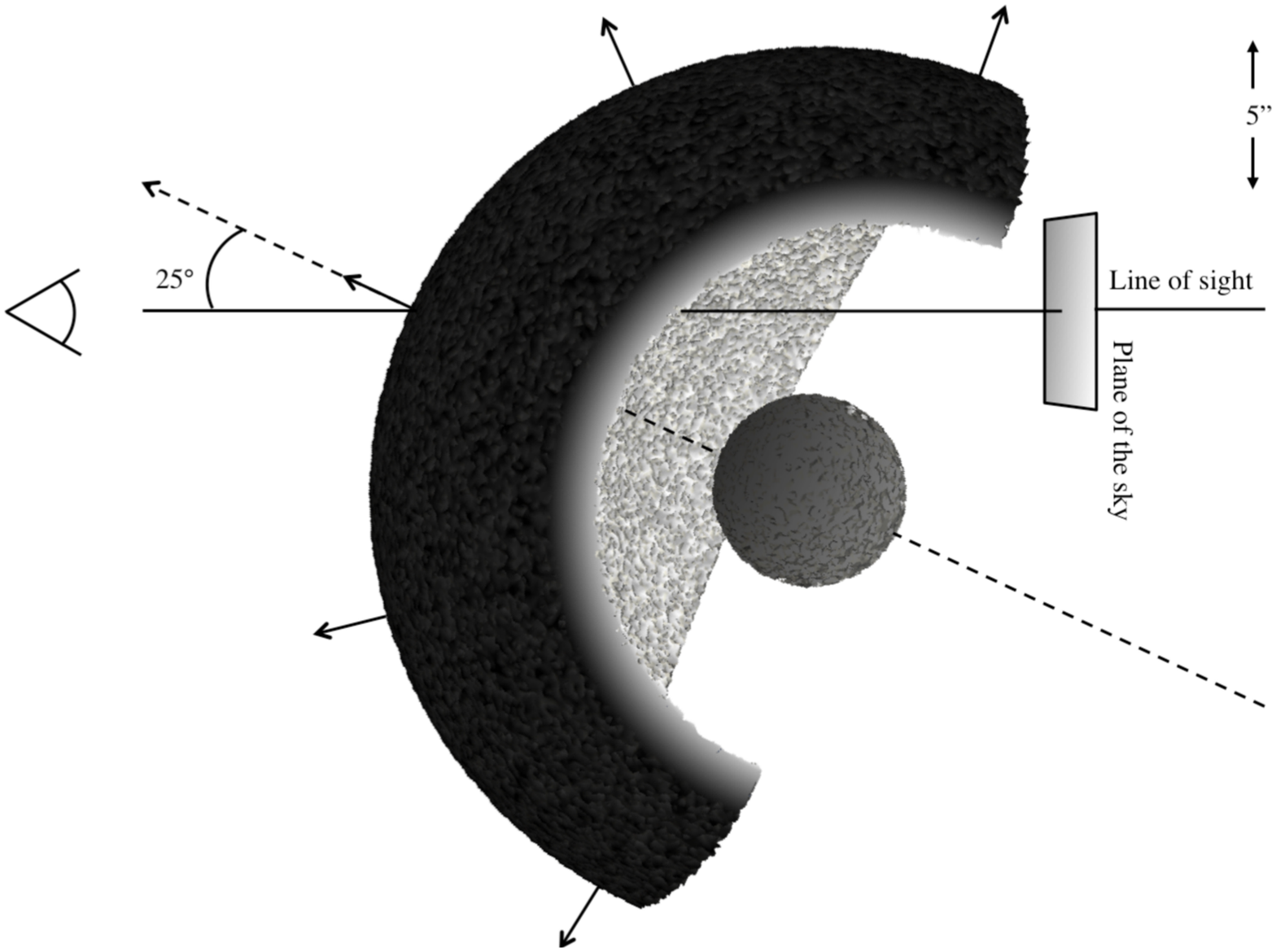}
         \caption{A 3D view of the model described in the text. The colours represent the magnitude of the velocity field, increasing from the inside (light grey) of the bow-shock component to the outside (dark grey). The direction of the velocity field is indicated with arrows and the motion of the Herbig-Haro object is indicated with a dashed line. The hot component, producing the bump spectral feature is located at the centre of the model and the inclination with respect to the line of sight is 25\adeg. A cut has been made through the top of the model for clarity.}
 \label{fig:model}
\end{figure}
The size of the emitting region is set to 7\asec\ \citep{Neufeld:2006fk}, the velocity is set to \mbox{\vlsr~=~--7~\kmpers}, and the velocity dispersion to 5~\kmpers. The bow-shock is again represented by an expanding shell at constant density and temperature. In the model presented here, however, the rear side of the sphere and the central blackbody source \citep{Bjerkeli:2011qy}  are removed, and the shell thickness has been decreased by a factor of two, i.e. $\Delta s$~=~ 2.5\texpo{15}~cm (see Table~\ref{table:limemodel} and Fig.~\ref{fig:model} for details). 500 000 grid points are used in the calculation.
\begin{table}[t]  
\flushleft
\caption{Model parameters}              
\label{table:limemodel}     
\resizebox{\hsize}{!}{
\begin{tabular}{l l l}          
  \hline\hline                        
  \noalign{\smallskip}
  \noalign{\smallskip}
  & \textit{Bow-shock component:} & \textit{Bump component:}\\
  \noalign{\smallskip}
 Kinetic temperature & 180~K & 800~K \\
 \htva\ density & \expo{5}~\cmthree & \expo{4}~\cmthree \\
  CO abundance & \expo{-4} & \expo{-4}  \\
  Velocity of component & +2.4~\kmpers & $-7.0$~\kmpers \\
   Source radius &$R_{\rm max}$~=~15\asec & $R_{\rm max}$~=~3\asec\\

   Shell thickness &$\Delta s / R_{\rm max}$ = 0.06 & - \\
  Velocity profile &$\upsilon(s)~=~20~s/\Delta s$~\kmpers & - \\ 
  Velocity dispersion & 1.5~\kmpers & 5.0~\kmpers  \\
  \noalign{\smallskip}
  \noalign{\smallskip}
  \hline                                            
  
\end{tabular}
}
\end{table}

\begin{figure}[ht]
   \centering
          \includegraphics[width=0.48\textwidth]{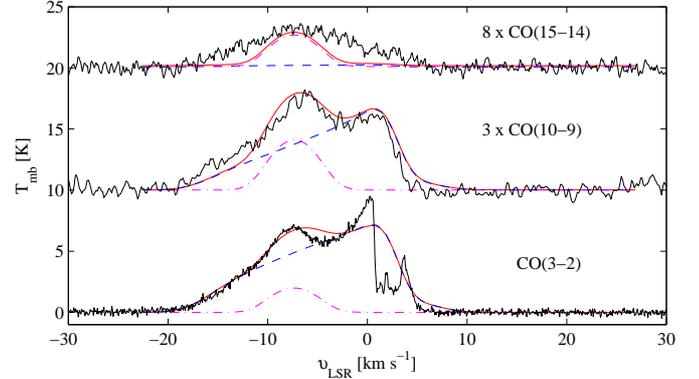}  
      \caption{The observed \cotretva, \cotionio, and \cofemtonfjorton\ spectra compared with the best-fit two-component model (red). The upper two lines have been shifted upwards and the foreground emission and absorption is not included in the model. The contributions from the bump and the expanding shell have been calculated separately and are indicated with dash-dotted purple and dashed blue lines, respectively.}
         \label{fig:limemodels}
   \end{figure}

In Fig.~\ref{fig:limemodels}, we present the \cotretva, \cotionio, and \cofemtonfjorton\ line profiles observed towards HH\,54, together with the model that fit the observations best. Given the increased number of free parameters, and because a full chi-square analysis is impracticable, we use by-eye comparison when fitting the lines. From Fig.~\ref{fig:limemodels}, it is clear that the observed lines, in essence, can be reproduced using this simplistic 3D model of an expanding bow-shock with a slightly under-dense and hot component. The spatial shift, due to the inclination (\about5\asec), between the centroids of the modelled low-$J$ and high-$J$ CO lines, can to a great extent explain the observed displacement with increasing $J_{\rm{up}}$ (Fig.~\ref{fig:gaussianfits}). We were not able to reproduce the line shapes, when using a high density and low temperature for the bump component. Why the \cofemtonfjorton\ line (where the line strength is completely accounted for by the hot component in the model) has a similar line width as the low-$J$ CO lines is, however, at present not fully understood. The inferred CO column density is 1\texpo{16}~\cmtwo, for the low-temperature component. The \htva\ column density of 1\texpo{20}~\cmtwo, is consistent with the column density obtained from a population diagram analysis using \spitzer\ \htva\ data \citep{Neufeld:2006fk}. From inspection of these diagrams, the S(0) line and the S(1) line seem to trace gas at lower excitation \citep[as in the case of \vlarhooph,][]{Bjerkeli:2012fk}. Taking only these two lines into account (where $E_{\rm{up}}/\rm{k} < 1020~K$), and assuming an ortho-to-para ratio of 3, we estimate the \htva\ column density in the region to lie in the range 0.5 -- 1\texpo{20}~\cmtwo. Taking the uncertainties of the above values into account, the assumed CO/\htva\ ratio of 1\texpo{-4}, for the low-temperature component, is estimated to be correct to within a factor of 4.
\section{Conclusions}
The FIR spectrum towards HH\,54 has been observed successfully. CO lines from $J_{\rm{up}}$~=~4 to 22, 9 \htvao\ lines, and the \oishort\ and \oilong\ lines, are detected. 
The CO transitions observed with SPIRE and PACS show a coherent displacement in the emission maximum with increasing $J$. The gradient goes in the direction of the source IRAS 12500 -- 7658, which has been proposed to be the exciting source of the HH\,54 flow. We interpret this as an under-dense, hot component present upstream of the HH\,54 shock wave. The peak positions of the high-$J$ CO lines are spatially coincident with the peaks of [O\,{\sc{i}}], [Ne\,{\sc ii}], [Fe\,{\sc ii}], [S\,{\sc i}], and [Si\,{\sc ii}]. Non-LTE radiative transfer modelling, taking the full 3D geometry into account can explain the observed CO line profile shapes and the spatial shift between the various CO lines. The CO/\htva\ ratio for the low-temperature gas is estimated \mbox{at \about\expo{-4}.} 

\begin{acknowledgements}
Per Bjerkeli appreciate the support from the Swedish research council (VR) through the contract 637-2013-472. 
 The Swedish authors also acknowledge the support from the Swedish National Space Board (SNSB). 
\end{acknowledgements}

\bibliographystyle{aa} 
\bibliography{papers}
\end{document}